# Strong enhancement of direct magnetoelectric effect in strained ferroelectric-ferromagnetic thin-film heterostructures

N. A. Pertsev,[1,2] H. Kohlstedt,[2] and B. Dkhil[3]

[1]*A. F. Ioffe Physico-Technical Institute, Russian Academy of Sciences, 194021 St. Petersburg, Russia*

[2]*Institut für Festkörperforschung, Forschungszentrum Jülich, D-52425 Jülich, Germany*

[3]*Laboratoire Structures, Propriétés et Modélisation des Solides, Ecole Centrale Paris, UMR-CNRS 8580, F-92295 Châtenay-Malabry, France*

The direct magnetoelectric (ME) effect resulting from the polarization changes induced in a ferroelectric film by the application of a magnetic field to a ferromagnetic substrate is described using the nonlinear thermodynamic theory. It is shown that the ME response strongly depends on the initial strain state of the film. The ME *polarization* coefficient of the heterostructures involving Terfenol-D substrates and compressively strained $Pb(Zr_{1-x}Ti_x)O_3$ (PZT) films, which stabilize in the out-of-plane polarization state, is found to be comparable to that of bulk PZT/Terfenol-D laminate composites ($\sim 5 \times 10^{-8}$ s/m). At the same time, the ME *voltage* coefficient reaches a giant value of 50 V cm$^{-1}$ Oe$^{-1}$, which greatly exceeds the maximum observed static ME coefficients of bulk composites. This remarkable feature is explained by a favorable combination of considerable strain sensitivity of polarization and a low electric permittivity in compressively strained PZT films. The theory also predicts a further dramatic increase of ME coefficients at the strain-induced transitions between different ferroelectric phases.

The direct magnetoelectric (ME) effect is potentially useful for the magnetic field sensing, novel magnetic recording read heads, current measurement probes for high-power electric transmission systems, and for energy harvesting.[1-3] Since the ME response of single-phase magnetoelectric materials is too small for device applications,[4] intensive experimental and theoretical studies were focused on multiferroic composites combining ferroelectric and ferromagnetic materials.[5] It was found, in particular, that laminate composites involving piezoelectric $Pb(Zr_{1-x}Ti_x)O_3$ (PZT) and magnetostrictive Terfenol-D layers display very high ME voltage coefficients exceeding 1 V cm$^{-1}$ Oe$^{-1}$.[1] The replacement of a ceramic PZT layer in such composite by the $\langle 001 \rangle$-oriented $Pb(Mg_{1/3}Nb_{2/3})O_3$–$PbTiO_3$ (PMN-PT) single crystal increases the ME coefficient up to 10 V cm$^{-1}$ Oe$^{-1}$.[1] Interestingly, considerable ME effect was found recently even in industrially produced multilayer capacitors consisting of alternating layers based on $BaTiO_3$ and Ni.[3] In addition to bulk ME composites, thin-film multiferroic heterostructures were investigated, such as PZT-Pd-$Co_xPd_{1-x}$ multilayers[6] and $BaTiO_3$-$CoFe_2O_4$ columnar nanostructures[7] fabricated on passive substrates and PZT films grown on the ferromagnetic $La_{1.2}Sr_{1.8}Mn_2O_7$ crystal.[8]

The theoretical studies of ME effects in multiferroic composites and heterostructures were predominantly based on *linear* constitutive equations.[5,9-12] The linear theory predicted, in particular, that the ME voltage coefficient of PZT/Terfenol-D multilayers with ideal interfacial coupling may be as large as 5 V cm$^{-1}$ Oe$^{-1}$.[9] The influence of imperfect interfacial bonding on the ME responses of multiferroic laminates was later modeled phenomenologically.[10] Using a linear theory, Nan *et al.* also evaluated the direct ME effect in the thin-film ferroelectric-ferromagnetic heterostructures fabricated on thick passive substrates.[11] These authors did not study, however, the expected dependence of the magnetic-field-induced polarization on the lattice strains created in the ferroelectric phase by a dissimilar substrate. Moreover, the mechanical boundary conditions employed in their calculations correspond to the three-dimensional clamping of the heterostructure (all components of the strain tensor are fixed), whereas the substrate actually produces the two-dimensional clamping only (three strains and three stresses remain constant).[13] Accordingly, a finite ME response was incorrectly predicted for a ferromagnetic-ferroelectric bilayer clamped by a thick substrate.

Recently, Liu *et al.* attempted to calculate the magnetically induced polarization in BaTiO$_3$-CoFe$_2$O$_4$ heterostructures with the aid of the Landau-Ginzburg-Devonshire thermodynamic theory of ferroelectrics.[14,15] However, they used incorrect approach based on the minimization of the Gibbs free energy $G$, which does not describe the effect of *internal* mechanical stresses on the physical properties of such heterostructures. Indeed, the elastic energy contribution to $G$ is negative, which is valid only for material systems subjected to *external* stresses. The correct theoretical approach to the description of internally strained heterostructures, therefore, should be based on the minimization of the modified thermodynamic potential $\tilde{G}$ or the Helmholtz free energy $F$, where the elastic energy contribution is positive.[13,16]

In this paper, the direct magnetoelectric effect in hybrid ferroelectric-ferromagnetic heterostructures is described using a *nonlinear* thermodynamic theory. We focus on material systems with the 2-2 connectivity, which combine a thin ferroelectric film with a thick ferromagnetic substrate in order to maximize this effect. The influence of lattice strains on the polarization and electric permittivity of a ferroelectric film is explicitly taken into account in the calculations. It is shown that the strain state of the ferroelectric film has a strong impact on the ME response of the studied multiferroic heterostructure.

Consider a single-crystalline ferroelectric film sandwiched between two continuous electrodes. The film is assumed to be epitaxially grown on a bottom electrode ensuring the (001) crystallographic orientation of the film lattice, as it happens in many ferroelectric heterostructures. The electrode-ferroelectric-electrode trilayer should be fabricated on a thick ferromagnetic substrate, which may require the deposition of additional buffer layer (Fig. 1). The initial strain state of an epitaxial film is defined by the lattice matching to the bottom electrode, which may be strained itself due to the mechanical interaction with the substrate. Restricting our analysis by the films of perovskite ferroelectrics grown on the (001)-oriented cubic or tetragonal bottom electrodes, we can introduce the relations $u_{11} = u_{22} = u_m$ and $u_{12} = 0$ for the in-plane strains imposed on the prototypic cubic state of a ferroelectric material. The misfit strain $u_m$ is given by the formula $u_m = (b^* - a_0)/a_0$, where $b^*$ is the effective in-plane lattice parameter of the bottom electrode, and $a_0$ is the lattice constant of a stress-free prototypic phase.[13]

3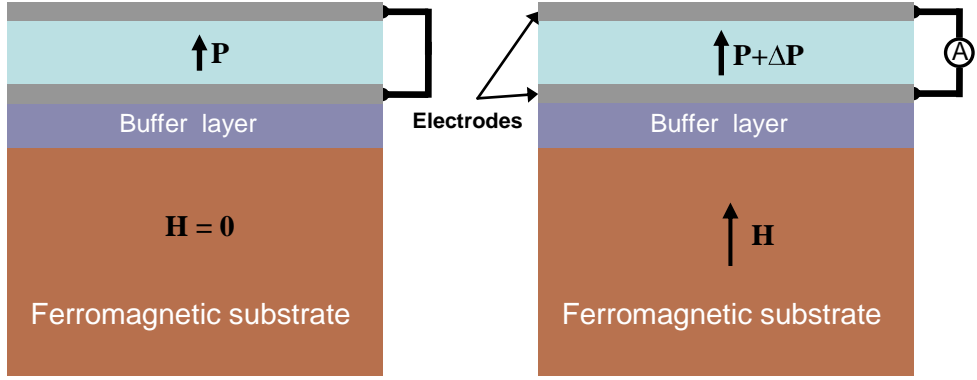

**FIG. 1.** Schematic representation of a hybrid material system involving a thin ferroelectric film deposited on a thick ferromagnetic substrate. The film is sandwiched between two continuous electrodes enabling the measurements of magnetoelectric coefficients.

When an external magnetic field **H** is applied to the heterostructure, macroscopic deformations $u_{ij}^s(\mathbf{H})$ appear in a ferromagnetic substrate due to the magnetostriction. In the material systems with perfect mechanical coupling at the interfaces, the substrate deformations change the in-plane film strains as $u_{11} = u_m + u_{11}^s(\mathbf{H})$, $u_{22} = u_m + u_{22}^s(\mathbf{H})$, and $u_{12} = u_{12}^s(\mathbf{H})$. Then the ME polarization coefficients $\alpha_{ij}$ can be calculated from the relation

$$\alpha_{ij} = \frac{\partial \langle P_i \rangle}{\partial u_{11}}\frac{\partial u_{11}^s}{\partial H_j} + \frac{\partial \langle P_i \rangle}{\partial u_{22}}\frac{\partial u_{22}^s}{\partial H_j} + \frac{\partial \langle P_i \rangle}{\partial u_{12}}\frac{\partial u_{12}^s}{\partial H_j}, \quad (1)$$

where $\langle P_i \rangle$ are the mean values of the polarization components $P_i$ ($i = 1,2,3$) in the ferroelectric film. In the symmetric case ($u_{11}^s = u_{22}^s$, $u_{12}^s = 0$), which will be analyzed below, Eq. (1) reduces to

$$\alpha_{i3} = \frac{\partial \langle P_i \rangle}{\partial u_m}\frac{\partial u_{11}^s}{\partial H_3}, \quad (2)$$

where the derivatives $\partial \langle P_i \rangle / \partial u_m$ correspond to the dependence of mean polarization components on the isotropic biaxial strain $u_m = u_{11} = u_{22}$.

Equation (2) describes the situation, where the magnetic field **H** is applied along the $x_3$ axis orthogonal to the interface. In addition, to ensure the equality $u_{11}^s = u_{22}^s$ and the absence of field-induced shear deformation $u_{12}^s$, this axis should represent a symmetry axis of the substrate at least of the fourth-order. This condition is satisfied, in particular, for polycrystalline Terfenol-D, a highly magnetostrictive alloy of iron and terbium and dysprosium, which represents a suitable substrate material strongly enhancing the direct ME effect. The field-induced deformations $u_{ii}^s$ of Terfenol-D vary nonlinearly with the magnetic field intensity, with the maximum longitudinal deformation $u_{33}^s$ being of the order of 0.1%.[17] Although the deformation $u_{33}^s(H_3)$ follows at small intensities the quadratic field dependence characteristic of magnetostriction and eventually saturates at high magnetic fields, it shows a quasi-linear variation in the range of intermediate field intensities. For our purposes, it is sufficient to characterize Terfenol-D by an effective piezomagnetic coefficient $\partial u_{33}^s / \partial H_3 \approx 10^{-8}$ m/A displayed at $H_3 < 2$ kOe.[17] Then, evaluating the transverse deformations from the condition $u_{11}^s = u_{22}^s = -0.5 u_{33}^s$ ensuring the volume conservation, we obtain $\partial u_{11}^s / \partial H_3 \approx -5 \times 10^{-9}$ m/A for the second factor in Eq. (2).

Proceed now to the calculation of the derivatives $\partial \langle P_i \rangle / \partial u_m$ defining the first factor in Eq. (2). These derivatives depend on the strain sensitivities $S_i = \partial |P_i|/\partial u_m$ of polarization components and on the domain structure existing in an epitaxial ferroelectric film. In order to maximize the discussed ME effect, the film should be poled prior to the measurements by an electric field sufficient to remove antiparallel 180° domains. Owing to the screening effect of metallic electrodes, thus formed homogeneous polarization state will remain stable after switching off the poling field.[18] Therefore, we shall assume that the film acquires a single-domain state so that the magnitude of $\partial \langle P_i \rangle / \partial u_m$ becomes equal to the strain sensitivity $S_i$.[19] The latter can be calculated using the nonlinear thermodynamic theory of epitaxial ferroelectric films through the dependences of polarization components on the misfit strain $u_m$.[20] For the tetragonal $c$ phase ($P_1 = P_2 = 0$, $P_3 \neq 0$) formed in compressively strained films ($u_m < 0$), the out-of-plane polarization $P_3$ can be calculated analytically. In the $P^6$ approximation valid for PbTiO$_3$ and PZT films, for example, the calculation gives the following relation:

$$P_3^2 = -\frac{a_{33}^*}{3a_{111}} + \sqrt{\frac{a_{33}^{*2} - 3a_1 a_{111}}{9a_{111}^2} + \frac{2Q_{12}}{3a_{111}(s_{11}+s_{12})}u_m}, \quad (3)$$

where $a_{33}^* = a_{11} + Q_{12}^2/(s_{11}+s_{12})$, $a_1$, $a_{11}$, and $a_{111}$ are the dielectric stiffness and higher-order stiffness coefficients at constant stress, $Q_{12}$ is the electrostrictive constant, and $s_{ln}$ are the film elastic compliances at constant polarization. It should be noted that Eq. (3) does not take into account the suppression of $P_3$ by the depolarizing field since it should be negligible in the discussed relatively thick films sandwiched between continuous electrodes, which are short-circuited before the ME measurements. The influence of depolarizing field on the ME polarization coefficient $\alpha_{33} = \partial P_3 / \partial H_3$ can be neglected as well, because $\alpha_{33}$ is evaluated by measuring an electric current flowing in the external circuit (see Fig. **1**).

Using Eq. (3), we obtain the strain sensitivity $S_3 = \partial |P_3|/\partial u_m$ of polarization in the $c$ phase as

$$S_3 = \frac{Q_{12}}{2(s_{11}+s_{12})|P_3|(a_{33}^* + 3a_{111}P_3^2)}. \quad (4)$$

It can be seen that $S_3$ depends on the film polarization $P_3(u_m)$ and, therefore, is not a constant quantity. Substituting into Eqs. (3) and (4) the involved material parameters known for PZT solid solutions,[21,22] we can quantify the strain sensitivity of polarization in PZT films. The calculation shows that, at a representative large compressive strain of $u_m = -16\times10^{-3}$, the sensitivity at room temperature amounts to $S_3 \approx -6.4$ C/m$^2$ in PZT 20/80 films and to $S_3 \approx -9.6$ C/m$^2$ in PZT 40/60 ones. When the misfit strain approaches a critical value $u_m^*$ above which the $c$ phase becomes less stable than the $r$ phase,[22] the magnitude of strain sensitivity increases by almost two times (see Fig. **2**), with $S_3 \approx -11.4$ C/m$^2$ in the PZT 20/80 film ($u_m^* \cong +2.43\times10^{-3}$) and $S_3 \approx -16.8$ C/m$^2$ in the PZT 40/60 one ($u_m^* \cong -2.95\times10^{-3}$).

Combining the calculated values of the strain sensitivity in the tetragonal $c$ phase with the effective piezomagnetic coefficient $\partial u_{11}^s / \partial H_3 \approx -5\times10^{-9}$ m/A given above, we find the ME polarization coefficient of PZT/Terfenol-D heterostructures to be in the range $\alpha_{33} \approx (3\div6)\times10^{-8}$ s/m at the Zr content $x = 20\%$ and $\alpha_{33} \approx (5\div8)\times10^{-8}$ s/m at $x = 40\%$. These values are close to the maximum ME response $\alpha_{33} \approx 6.5\times10^{-8}$ s/m measured in PZT/Terfenol-D bulk laminate composites,[1] which indicates that our calculations correctly predict the order of magnitude of the direct ME effect.

The ME voltage coefficient $\alpha_{E33} = \partial E_3 / \partial H_3$, which defines the magnetically induced output voltage under open-circuit conditions, can be calculated from $\alpha_{33}$ via the relation $\alpha_{E33} = \alpha_{33}/(\varepsilon_0 \varepsilon_{33})$, where $\varepsilon_{33}$ is the relative out-of-plane permittivity of the ferroelectric film, and $\varepsilon_0$ is the permittivity of the vacuum. For epitaxial PZT films, the dielectric response was calculated as a function of the misfit strain in our earlier paper.[22] At $u_m = -16 \times 10^{-3}$, the permittivity $\varepsilon_{33}$ is about 60 in the PZT 20/80 film and about 80 in the PZT40/60 one, which gives $\alpha_{E33} \approx 50$ V cm$^{-1}$ Oe$^{-1}$ for both compositions. Since the film dielectric response increases significantly as $u_m$ tends to $u_m^*$, the ME voltage coefficient decreases down to about 40 V cm$^{-1}$ Oe$^{-1}$ near $u_m^*$ (see Fig. 3).

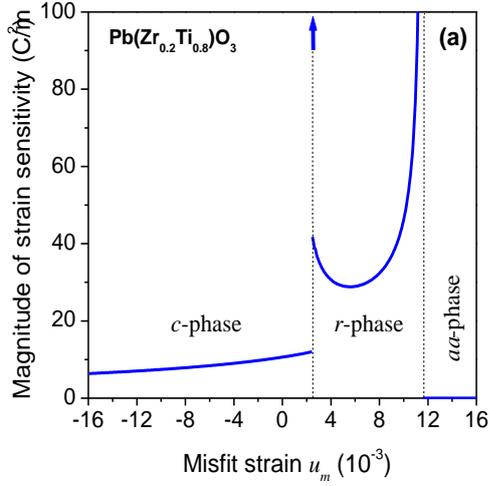

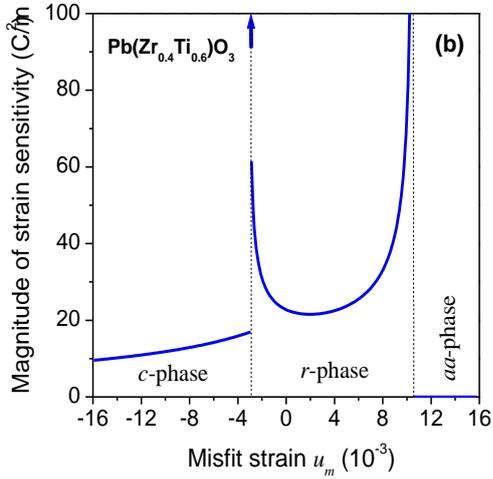

**FIG. 2.** Magnitude of the strain sensitivity $S_3$ of the out-of-plane polarization $P_3$ calculated for single-domain epitaxial PZT 20/80 (a) and 40/60 (b) films subjected to isotropic biaxial in-plane strain $u_m$. The arrow indicates drastic increase of the sensitivity at strain changes inducing the $r \rightarrow c$ phase transition.

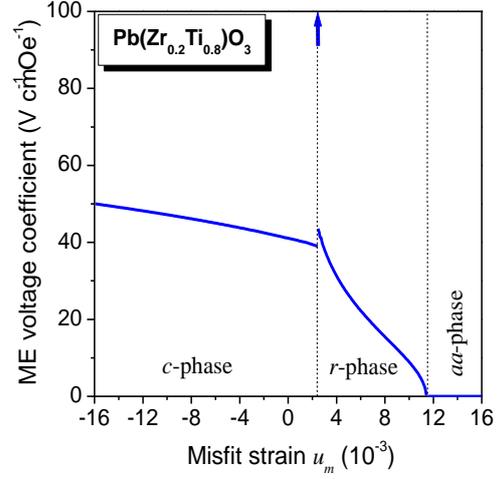

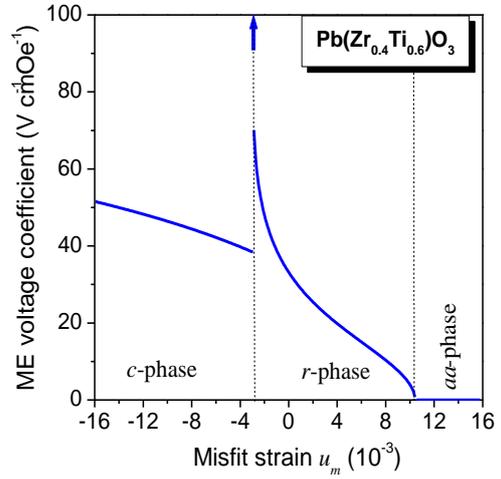

**FIG. 3.** Magnetoelectric voltage coefficient $\alpha_{E33}$ calculated for a hybrid heterostructure involving the Terfenol-D substrate and the single-domain epitaxial PZT 20/80 (a) or 40/60 (b) film subjected to isotropic biaxial in-plane strain $u_m$. The effective piezomagnetic coefficient $\partial u_{11}^s / \partial H_3$ of Terfenol-D is taken to be equal to $-5 \times 10^{-9}$ m/A. The arrow indicates drastic increase of $\alpha_{E33}$ at strain changes inducing the $r \rightarrow c$ phase transition.

Nevertheless, the magnitude of $\alpha_{E33}$ remains much larger than the maximum static ME coefficient displayed by the PZT/Terfenol-D bulk laminate composite ($\alpha_{E33} < 5$ V cm$^{-1}$ Oe$^{-1}$) and even by the PMN-PT/Terfenol-D one ($\alpha_{E33} \approx 10$ V cm$^{-1}$ Oe$^{-1}$).[1] This remarkable feature is explained by the fact that the out-of-plane permittivity of epitaxial PZT films stabilized in the tetragonal $c$ phase ($\varepsilon_{33} < 200$) is much lower than the permittivities of PZT ceramic

($\varepsilon_{33} \approx$ 1250) and PMN-PT single crystal ($\varepsilon_{33} \approx$ 4300) employed in the experimental investigations.

When the ferroelectric film in a hybrid heterostructure stabilizes in the monoclinic *r* phase ($u_m^* < u_m < u_m^{**}$), all three polarization components $P_i$ become different from zero so that the direct ME effect acquires new features. Using the dependences $P_i(u_m)$ reported for PZT films,[22] we calculated their strain sensitivities $S_3$ as a function of the misfit strain. Figure 2 shows variations of the sensitivity $S_3$ of the out-of-plane polarization in PZT 20/80 and 40/60 films.[23] It can be seen that $S_3$ varies nonmonotonically within the stability range of the *r* phase, being larger than in the *c* phase and rising steeply near the critical strains $u_m^*$ and $u_m^{**}$ limiting this range. Accordingly, the ME polarization coefficient $\alpha_{33}$ should also increase dramatically near the strain-induced phase transitions in the ferroelectric film. In contrast, the ME voltage coefficient $\alpha_{E33}$ is large only near the *r*→*c* phase transition (see Fig. 3), because enormous increase of the permittivity $\varepsilon_{33}$ at the second-order *r*→*aa* phase transition[22] makes the output voltage negligible.

It should be emphasized that the ME responses described above correspond to weak magnetic fields which do not change the phase state of a ferroelectric film during the ME measurement. If the applied magnetic field **H** is high enough to induce a phase transition in the film, the measured ME coefficient becomes strongly dependent on the magnitude of **H**. In particular, when the *r*→*c* transformation takes place, the polarization change $\Delta P_3(H_3)$ becomes mainly defined by a jump of $P_3$ at this first-order phase transition. Hence $\Delta P_3$ only weakly depends on $H_3$ so that $\alpha_{33}$ decreases rapidly with increasing field magnitude. Nevertheless, the calculation shows that the strain sensitivity $S_3$ of the PZT 20/80 and 40/60 films may exceed 100 C/m$^2$ (Fig. 2), which results in the ME polarization coefficient $\alpha_{33} > 5 \times 10^{-7}$ s/m and the ME voltage coefficient $\alpha_{E33} > 100$ V cm$^{-1}$ Oe$^{-1}$, as indicated in Fig. 3.

In summary, we studied theoretically the direct magnetoelectric effect in hybrid ferroelectric-ferromagnetic heterostructures. To that end, the strain sensitivity of polarization in PZT thin films was calculated as a function of the misfit strain in the epitaxial system. Taking into account the effective piezomagnetic coefficient of ferromagnetic substrate, we evaluated ME coefficients of PZT/Terfenol-D heterostructures. The ME voltage coefficient, which represents the figure of merit characterizing the performance of material system for magnetic-field sensors, was found to be enormously high for hybrids involving compressively strained ferroelectric films. In similarity with the converse ME effect in ferromagnetic-ferroelectric heterostructures,[24] the direct ME response may further increase dramatically at the strain-induced phase transitions in a thin film.

The financial support of the Deutsche Forschungsgemeinschaft is gratefully acknowledged.

---